\documentclass[fleqn,twoside,twocolumn,nofootinbib,showkeys]{revtex4} 
\usepackage[nocpr]{ujp} 

\begin{document}
\title[Calculation of Equilibrium Constant]
{CALCULATION OF EQUILIBRIUM CONSTANT\\ FOR DIMERIZATION OF HEAVY
WATER MOLECULES\\
IN SATURATED VAPOR}%
\author{L.A.~Bulavin}
\affiliation{Taras Shevchenko National University of Kyiv, Faculty of Physics}
\address{4, Academician Glushkov Ave., Kyiv 03127, Ukraine}
\author{S.V.~Khrapatyi}
\affiliation{Taras Shevchenko National University of Kyiv, Faculty of Physics}
\address{4, Academician Glushkov Ave., Kyiv 03127, Ukraine}
\author{V.M.~Makhlaichuk\,}%
\affiliation{I.I.~Mechnikov National University of Odessa}%
\address{2, Dvoryans'ka Str., Odesa 65026, Ukraine}%
\email{interaktiv@ukr.net}

\udk{???} \pacs{36.40.-c, 61.20.Ja} \razd{\secix}

\autorcol{L.A.\hspace*{0.7mm}Bulavin, S.V.\hspace*{0.7mm}Khrapatyi,
V.M.\hspace*{0.7mm}Makhlaichuk}

\setcounter{page}{263}%

\begin{abstract}
The magnitude and the temperature dependence of the equilibrium
constant of dimerization of heavy water molecules in saturated vapor
in terms of the second virial coefficient of the equation of state
have been determined.\,\,An expression is found for the equilibrium
dimerization constant of water vapor molecules, which contains terms
involving the monomer--monomer, monomer--dimer, and dimer--dimer
interaction.\,\,The obtained results are compared with experimental
data.\,\,The equilibrium constant of dimerization in heavy water
vapor is shown to exceed that in light water vapor within the whole
temperature interval.
\end{abstract}
\keywords{dimerization constant, heavy water.}

\maketitle

\section{Introduction}

The unusual properties of water (H$_{2}$O) have been known since
ancient times \cite{1,2,3}.\,\,As a rule, they are explained by the
existence of hydrogen bonds that arise between water molecules and
result in the formation of molecular complexes, such as dimers,
trimers, and so forth \cite{4,5,6}.\,\,The discovery of heavy water
(D$_{2}$O) and its further study showed that the substitution of
hydrogen by deuterium results in substantial changes of properties
in comparison with light water.\,\,For instance, the ternary point
temperature for D$_{2}$O is by 3~K higher than the corresponding
parameter for H$_{2}$O, whereas its critical temperature, on the
contrary, is by 4~K lower.\,\,The volatility of heavy water is lower
than that of light water.\,\,Heavy water is more
hygroscopic.\,\,Even the 30\% solution of heavy water in light one
is toxic and leads to the death of live organisms.\,\,When studying
the properties of the vapors of light and heavy waters, it is rather
successful to apply the virial equation of state, which is usually
confined to the second virial coef\-fi\-ci\-ent.\,\,The values of
second virial coefficient found experimentally for light and heavy
waters are different by a factor of $2$$\div$$3$ in the whole
temperature interval \cite{8}.\,\,One should expect that this fact
can considerably influence the value of equilibrium constant for the
molecular dimerization in heavy water \mbox{vapor.}\looseness=1

Note that the differences between H$_{2}$O and D$_{2}$O manifest
themselves not only in thermodynamic but also kinetic
properties.\,\,In particular, the kinematic viscosity along the
saturation curve of heavy water is about 25\% higher than the
kinematic viscosity of light water in the whole temperature interval
\cite{8}.\,\,This work aimed at calculating the dimerization degree
in saturated D$_{2}$O vapor with the help of the second virial
coefficient in the equation of state.

\section{Determination of the Dimerization\\ Constant for Molecules in Saturated
Vapor}

It is well known that the equilibrium properties of the dimerization
process ($m+m\Leftrightarrow d$) are described by chemical
thermodynamic methods.\,\,According to the latter, the chemical
potentials of monomers, $\mu_{m}$, and dimers, $\mu_{d}$, satisfy
the equality
\begin{equation}
\mu_{d}=2\mu_{m}.%
\end{equation}
At the same time, they are functions of the corresponding
concentrations.\,\,Therefore, Eq.\,(1) is actually an equation for
the indicated concentrations.\,\,The molar concentrations of water
monomers and water molecules united in dimers are defined as
\begin{equation}
c_{m}=n_{m}/n_{0},~~~~c_{d}=2n_{d}/n_{0},
\end{equation}
respectively, where
\begin{equation}
n_{0}=n_{m}+2n_{d}=\dfrac{N_{w}}{V}%
\end{equation}
is the initial density of water monomers in non-dimerized
vapor.\,\,According to the concentration definitions for monomers
and dimers (2), the condition of their normalization looks like
\begin{equation}
c_{m}+c_{d}=1.
\end{equation}
It was shown in work \cite{9} that, at small deviations of saturated
vapor
from the ideality, the concentration of dimers is determined as follows:
\begin{equation}
c_{d}\approx\zeta+...,
\end{equation}
where $\zeta=2n_{0}TK_{p}(T)$, and $K_{p}(T)$ is the dimerization
constant.\,\,In the general case, the chemical potentials of
components in a mixture of monomers and dimers contain additional
contributions associated with the interaction between particles,
which look like
\begin{equation}
\mu_{i}=\mu_{i}^{\rm (id)}+\mu_{i}^{\rm (ex)},
\end{equation}
where $i=m,d$.\,\,If the dimerization constant and, accordingly,
$\zeta =2n_{0}TK_{P}(T)$ are unknown, the combination of Eq.\,(1)
with the equation of state
\begin{equation}
P=n_{0}T(1+n_{0}B_{\mathrm{exp}}(T)+...)
\end{equation}
allows one to obtain an explicit expression for $K_{p}(T)$.\,\,In
work \cite{9}, a relation between the second virial coefficient
$B_{\mathrm{exp}}(T)$ in the equation of state, the dimerization
constant $K_{p}(T)$, and the parameters of the intermolecular
interaction in partially dimerized water vapor was
established.\,\,In the linear approximation in the concentration
$c_{d}$, we have
\begin{equation}
\begin{array}{l}
\displaystyle \zeta = \zeta_0, \\[2mm]
\displaystyle     \zeta_0 =\\
\displaystyle =
     \frac{B_{\mathrm{exp}}(T)-v_0^{(m)}+a_{11} / T}{p_1 (\frac{1}{2} v_0^{(d)}
     - \frac{3}{2} v_0^{(m)} - (a_{12}-2a_{11}) / T - 1 / (2 n_0))},
\end{array}\!\!\!\!\!\!\!\!\!\!\!\!\!\!\!\!\!\!\!\!\!\!\!\!
\end{equation}
where
\begin{equation}
p_{1}=1+2n_{0}(v_{0}^{(m)}-(a_{11}-\frac{1}{2}a_{12})/T),
\end{equation}
$v_{0}^{(m)}$ and $v_{0}^{(d)}$ are the excluded volumes of a
monomer and a dimer, respectively; and $a_{11}$ and $a_{12}$ are the
parameters of the van der Waals equation of state for a gas mixture,
which describes the excess pressure induced by the monomer--monomer
and monomer--dimer attraction forces, respectively.\,\,In the
quadratic approximation in the concentration $c_{d},$ we obtain
\begin{equation}
\zeta=\zeta_{0}+h\zeta_{0}^{2}+...,
\end{equation}
where%
\[
h = p_2 + 2p_1 \,\times
\]\vspace*{-7mm}
\[
\times\left(\!\!1\! -\! \frac{3 v_0^{(m)} - v_0^{(d)} - (7a_{11} -
4a_{12} + a_{22}) /\!T }{4 \left(-1 / n_0 + v_0^{(d)} - 3v_0^{(m)} -
\left(a_{12} - 2a_{11}\right)\right) /\! T }\!\!\right)\!\!,
\]\vspace*{-7mm}
\[
p_2 = 2n_0 (0{}.25 v_0^{(d)}\! + \!v_0^{(m)} - (a_{11}\! - \!a_{12}
\!+\! 0{}.25a_{22}) / T).
\]
The linear approximation in the concentration $c_{d}$ contains the
contributions that involve for only the monomer--monomer and
monomer--dimer attraction (the parameters $a_{11}$ and $a_{12}$ of
the equation of state in Eq.\,(8)).\,\,At the same time, the
quadratic approximation also includes contributions which are a
consequence of the dimer--dimer interaction (the parameter $a_{22}$
in the equation of state).\,\,For the saturated vapor of light
water, the account for those interactions is crucial \cite{9}.

\section{Calculation of the Dimerization Constant}

In order to determine the quantity $\zeta=2n_{0}TK_{p}(T)$ or,
equivalently, $K_{p}(T)$, we need to know the experimental values of
second virial coefficient $B_{\mathrm{exp}}(T)$, excluded volumes
$v_{0}^{(i)}$ ($i=m,d$), and gravitation constants $a_{mn}$
($m,n=1,2$) in the van der Waals equation.\,\,The value of second
virial coefficient for the saturated vapor of heavy water was
calculated proceeding from the experimental data on the pressure,
density, and temperature at the saturation curve \cite{8}.\,\,The
quantities $v_{0}^{(i)}$ ($i=m,d$) and $a_{mn}$ ($m,n=1,2$) are
connected with the behavior of intermolecular interaction
potentials.\,\,Let us take into account that water monomers and
dimers are permanently rotate in the gaseous state, so that the
microscopic potentials determining the interaction between water
molecules and dimers become effectively averaged
\cite{10,11,12,13}.\,\,A detailed discussion of the potential
averaging over the monomer and dimer orientations can be found in
works \cite{14,15}.

For the averaged potentials of interaction between water monomers and dimers,
let us take advantage of the Sutherland potential
\begin{equation}
U_{ij}(r_{12})=%
\begin{cases}
\infty, & r_{12}<\sigma_{ij},\\
-\varepsilon_{ij}(\frac{\sigma_{ij}}{r_{12}})^{6}, & r_{12}>\sigma_{ij},
\end{cases}
\end{equation}
where $i,j=m,d$.\,\,In this case, the quantities $v_{0}^{(m)}$ and
$a_{11}$ can be found, by using the known procedure \cite{16}.\,\,As
a result, they equal
\[
v_{0}^{(m)}=\frac{16\pi}{3}r_{m}^{3},\ \ a_{11}=\varepsilon_{m}v_{0}^{(m)}.
\]
In all further calculations, we suppose that the excluded volumes of
monomers and dimers coincide with the four-fold volumes of hard
spheres with the radii $r_{m}=1.58${\AA} and $r_{d}=2.98${\AA},
respectively.\,\,In addition, we adopt that the averaged values of
interaction constant equal to those quoted in Table~1 (see work
\cite{14}).\,\,Those values are approximately four times larger than
the constant of dispersion interaction \cite{10,11,12,13}, because
the dipole moment of each water molecule is induced by both the
fluctuations of the electron density at neighbor water molecules and
changes in the orientations of the bare dipole moments of those
molecules (see works \cite{14,15}).

In order to find $v_{0}^{(d)}$ and $a_{22}$, the rotation of dimers
has to be taken into consideration.\,\,Therefore, the dimer radius
should be taken equal to the monomer diameter: $r_{d}=\sigma_{mm}$.
The average polarizability of a rotating dimer
$\alpha_{d}\approx2\alpha_{m}$.\,\,Therefore, $\varepsilon
_{d}\approx4\varepsilon_{m}$.\,\,Then
\[
v_{0}^{(d)}=\frac{16\pi}{3}r_{d}^{3}\Rightarrow8v_{0}^{(m)},~~~~ a_{22}%
=\varepsilon_{d}v_{0}^{(d)}\Rightarrow32\varepsilon_{m}v_{0}^{(m)}.
\]
Making allowance for dimer rotations, the interaction between a dimer and
a monomer is described by the parameters $r_{dm}=3r_{m}$ and $\varepsilon
_{dm}=2\varepsilon_{mm}$, which brings us to the formula
\[
a_{12}\approx\frac{27}{4}v_{0}^{(m)}\varepsilon_{mm}.
\]
The results of calculations are presented in Table~2.

\section{Discussion of the Obtained\\ Results and Conclusions}

\begin{figure}%
\vskip1mm
\includegraphics[width=7cm]{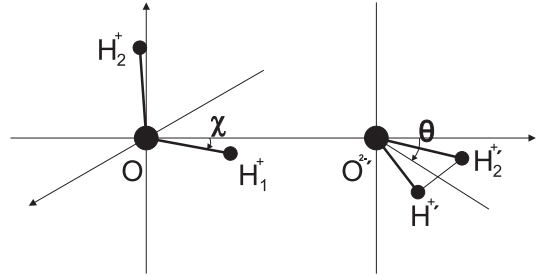}
\vskip-3mm\caption{Equilibrium configuration of a water dimer}
\end{figure}


\begin{table}[b!]
\noindent\caption{Averaged constant of interaction\\ between water
molecules, {\boldmath$\varepsilon_{m}$} ({\boldmath$k_{\mathrm{B}}$}
is the Boltzmann\\ constant, \boldmath$T_{c}$ is the critical
temperature)}\vskip3mm\tabcolsep10.4pt

\noindent{\footnotesize
\begin{tabular}{|c|c|c|c|c|}
 \hline%
 \multicolumn{1}{|c}{\rule{0pt}{5mm}\textit{T}}%
 & \multicolumn{1}{|c}{300 K}
 & \multicolumn{1}{|c}{400 K}
 & \multicolumn{1}{|c}{500 K}
 & \multicolumn{1}{|c|}{600 K}\\[2mm]%
\hline%
\rule{0pt}{5mm}$\varepsilon_m / K_\mathrm{B}T_c$ & 3.08 & 3.05 & 2.70 & 1.78 \\[2mm]%
\hline
\end{tabular}
}
\end{table}


\begin{table}[b!]
\vskip3mm \noindent\caption{Dimerization degree\\ and the dimerization
constant in the saturated\\ vapor of heavy
water}\vskip3mm\tabcolsep21.2pt

\noindent{\footnotesize
\begin{tabular}{|c|c|c|}
 \hline%
 \multicolumn{1}{|c}{\rule{0pt}{5mm}$T, K$}%
 & \multicolumn{1}{|c}{$c_d$ (D$_2$O)}
 & \multicolumn{1}{|c|}{$K_p(T)$ (D$_2$O)}\\[2mm]%
\hline%
\rule{0pt}{5mm}300 & 0.005 & 0.1580\\ 
325 & 0.016 & 0.1279 \\
350 & 0.034 & 0.0858 \\
375 & 0.064 & 0.0593 \\
400 & 0.102 & 0.0387 \\
425 & 0.149 & 0.0257 \\
450 & 0.202 & 0.0166 \\
475 & 0.251 & 0.0104 \\
500 & 0.295 & 0.0063 \\
525 & 0.343 & 0.0038 \\
550 & 0.373 & 0.0022 \\
575 & 0.439 & 0.0014 \\
600 & 0.552 & 0.0009 \\
625 & 0.886 & 0.0008 \\[2mm]%
\hline
\end{tabular}
}
\end{table}

\begin{figure}%
\vskip1mm
\includegraphics[width=\column]{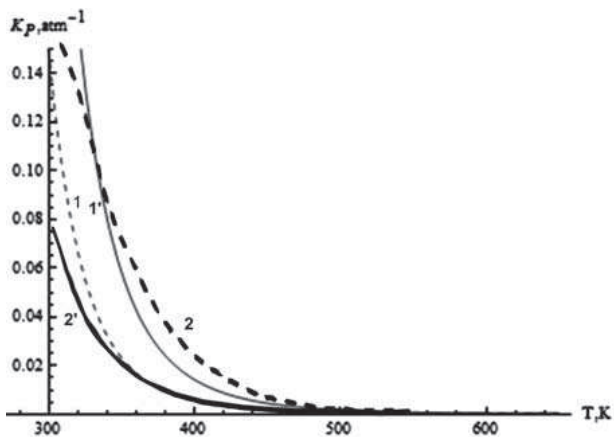}
\vskip-3mm\caption{Temperature dependences of the dimerization constant
for the saturated vapors of heavy (\textit{1}) and light
(\textit{1}$^{\prime}$) waters according to the results of work
\cite{19} and for heavy (\textit{2}) and light
(\textit{2}$^{\prime}$) waters according to the calculations by
formula~(10)}
\end{figure}


\begin{table}[t!]
\vskip5mm \noindent\caption{Rotational quanta of dimers (in
\textrm{cm}\boldmath$^{-1}$\\ units) for various intermolecular
interaction potentials}\vskip3mm\tabcolsep5.7pt

\noindent{\footnotesize
\begin{tabular}{|l|c|c|c|c|c|c|}
 \hline%
 \multicolumn{1}{|c}{\rule{0pt}{5mm}\raisebox{-3mm}[0cm][0cm]{Potential model}}%
 & \multicolumn{3}{|c}{H$_2$O}
 & \multicolumn{3}{|c|}{D$_2$O} \\[2mm]
\cline{2-7}
\rule{0pt}{5mm}&$Ox$&$Oy$&$Oz$&$Ox$&$Oy$&$Oz$\\[2mm]
\hline 
\rule{0pt}{5mm}~~~~~~~GSD & 0.21  & 8.65  & 0.21  & 0.18  & 4.32  & 0.21 \\
~~~~~~~SPC & 0.23  & 8.49  & 0.22  & 0.20  & 4.24  & 0.24 \\
~~~~~~~SPC/E   & 0.24  & 8.39  & 0.22  & 0.20  & 4.19  & 0.25 \\
~~~~~~~TIPS    & 0.24  & 9.42  & 0.22  & 0.20  & 4.71  & 0.25 \\
~~~~~~~TIP3P   & 0.24  & 9.44  & 0.22  & 0.20  & 4.72  & 0.25 \\
~~~~~~~SPCM    & 0.21  & 9.58  & 0.23  & 0.18  & 4.79  & 0.21 \\[2mm]%
\hline
\end{tabular}
}
\end{table}
\begin{table}[t!]
\vskip2mm  \noindent\caption{Frequencies (in \textrm{cm}\boldmath$^{-1}$\\
 units) of vibrations in H$_{2}$O and D$_{2}$O
dimers\\ for the SPC and TIPS potentials
\cite{18}}\vskip3mm\tabcolsep12.6pt

\noindent{\footnotesize
\begin{tabular}{|c|c|c|c|c|}
 \hline%
 \multicolumn{3}{|c}{\rule{0pt}{5mm}SPC}%
 & \multicolumn{2}{|c|}{TIPS} \\[2mm]
\hline
\rule{0pt}{5mm}&H$_2$O&D$_2$O&H$_2$O&D$_2$O\\[2mm]
\hline 
\rule{0pt}{5mm}$\omega_1$ & ~\,70.51  & ~\,49.86 & ~\,73.71 & ~\,52.12 \\
$\omega_2$ & 240.73 & 170.22 & 212.63 & 172.24 \\
$\omega_3$ & 246.00 & 219.58 & 243.59 & 201.72 \\
$\omega_4$ & 310.53 & 233.37 & 322.32 & 227.19 \\[2mm]%
\hline
\end{tabular}
}\vspace*{-4mm}
\end{table}

The analysis of the data quoted in Table~2 and work \cite{9}
demonstrates that the difference between the dimerization constants
of heavy and light water vapors is rather substantial.\,\,Depending
on the temperature, they differ from each other by a factor of 2 to
3.\,\,From the principal viewpoint, this is a result of the
difference between the character of thermal excitations in heavy and
light water dimers.\,\,Concerning the corresponding parameters of
dimer ground states, they are close to one another.\,\,At the same
time, the rotational quanta of heavy and light water dimers
$Q_{i}=\hbar^{2}/2I_{i}$, where $\hbar$ is Planck's constant, and
$I_{i}$ is the moment of inertia with respect to the $i$-th axis,
are different.\,\,In the case of the dimer configuration depicted in
Fig.~1, the rotational quanta of dimers for various potentials of
intermolecular interaction are indicated in Table~3.\,\,The
corresponding differences amount to 15--20\% for rotations about the
axes $x$ and $z$ and approximately 100\% for rotations about the
\mbox{axis $y$ \cite{17}.}

The energies of vibrational excitations are also
con\-si\-de\-rab\-ly different.\,\,The frequencies of small
vi\-bra\-tions in H$_{2}$O and D$_{2}$O dimers are compared in
\mbox{Table~4.}

In Fig.~2, the results of calculations of the dimerization constant
for light and heavy water vapors carried out on the basis of the
second virial coefficient in the equation of state are shown, as
well as the results of direct calculations of the dimerization
constant using the statistical physics methods by determining the
internal partition functions of monomers and dimers \cite{19}.\,\,It
is evident that the dimerization constants determined on the basis
of experimental values for the second virial coefficient correlate
well with the results of theoretical calculations of the
dimerization constants obtained in works \cite{20,21,22}.\,\,One can
see that the equilibrium constant of dimerization of heavy water
molecules substantially depends on the effects of the interaction
between monomers and dimers.\,\,It is owing to this interaction that
the dimerization of molecules takes place.\,\,On the basis of the
experimental values of second virial coefficient, we obtained the
value of dimerization constant.

Attention should be paid to the fact that the temperature
dependences of the dimerization constant for light and heavy waters,
which were calculated using different methods, have an opposite
relative arrangement.\,\,Moreover, at temperatures in a vicinity of
the ternary point, a considerable discrepancy is observed between
the values obtained for $K_{p}(T)$ by different
methods.\,\,Unfortunately, we cannot explain now the origin of this
difference, because the values of second virial coefficient in the
temperature interval 300--459$~\mathrm{K}$, which were used at
calculations, are not quite reliable.\,\,It is so because 1)~there
is a discrepancy in the determination of experimental values for the
parameters in the equation of state \cite{8,23}, and 2)~there is no
possibility to verify the values of $B_{\mathrm{exp}}(T)$ with the
help of experimental values obtained for the viscosity of heavy
water vapor.\,\,In addition, in the case of the direct calculation
of $K_{p}(T)$, the determination accuracy for vibrational
frequencies is directly connected with the choice of intermolecular
potentials.\,\,Unfortunately, it is difficult to specify, which of
the potentials used in the literature is the most
adequate.\,\,Moreover, intramolecular vibrational and rotational
modes are considered independent in work \cite{19}.\,\,However, at
the large values of rotational quantum number, the dimer parameters
considerably differ from their values in the ground state, which
should be accompanied by changes in the vibrational and rotational
modes.\,\,At temperatures higher than 400$~\mathrm{K}$, the relative
influence of indicated factors decreases, and the values of
equilibrium dimerization constant for saturated vapors of light and
heavy water molecules agree with the results of direct calculations
carried out in work \cite{19}.\,\,We intend to study those issues
elsewhere in more detail.\,\,\vskip3mm

\textit{ The authors express their sincere gratitude to
Prof.\,\,M.P.\,\,Ma\-lomuzh for the fruitful discussion of ideas and
results of this work.}

\rezume{Л.А.\,Булавін, С.В.\,Храпатий, В.М.\,Махлайчук}{РОЗРАХУНОК
КОНСТАНТИ\\ РІВНОВАГИ ДИМЕРІЗАЦІЇ МОЛЕКУЛ\\ НАСИЧЕНОЇ ПАРИ ВАЖКОЇ
ВОДИ} {Робота присвячена визначенню величини та температурної
залежності константи рівноваги  димеризації молекул насиченої пари
важкої води відповідно до другого віріального коефіцієнта  рівняння
стану. Знайдено вираз для константи рівноваги димеризації молекул
водяної пари, який містить доданки, що враховують взаємодію
мономер--мономер, мономер--димер і димер--димер. Проведено
порівняння отриманих результатів з експериментальними даними.
Показано, що у всій області температур константа рівноваги
димеризації пари важкої води перевищує константу рівноваги
димеризації молекул пари легкої води.}

\end{document}